\documentclass[aps,prl,twocolumn,superscriptaddress,showpacs]{revtex4}
\usepackage[latin1]{inputenc}
\usepackage{graphicx}
\usepackage{amsmath,amsfonts}

\begin{document}

\title{Theory of Plasmon-assisted Transmission of Entangled Photons}

\author{Esteban Moreno}
\email[Electronic address: ]{esteban.moreno@uam.es}
\affiliation{Laboratory for Electromagnetic Fields and Microwave
Electronics, Swiss Federal Institute of Technology, ETH-Zentrum,
Gloriastrasse 35, CH-8092 Zurich, Switzerland}

\author{F. J. García-Vidal}
\affiliation{Departamento de Física Teórica de la Materia
Condensada, Universidad Autónoma de Madrid, E-28049 Madrid, Spain}

\author{Daniel Erni}
\affiliation{Laboratory for Electromagnetic Fields and Microwave
Electronics, Swiss Federal Institute of Technology, ETH-Zentrum,
Gloriastrasse 35, CH-8092 Zurich, Switzerland}

\author{J. Ignacio Cirac}
\affiliation{Max Planck Institut für Quantenoptik, Hans-Kopfermann
Strasse 1, D-85748 Garching, Germany}

\author{L. Martín-Moreno}
\affiliation{Departamento de Física de la Materia Condensada,
Universidad de Zaragoza-CSIC, E-50009 Zaragoza, Spain}

\date{\today}

\begin{abstract}
The recent surface plasmon entanglement experiment [E. Altewischer
\emph{et al.}, Nature (London) \textbf{418}, 304 (2002)] is
theoretically analyzed. The entanglement preservation upon
transmission in the non-focused case is found to provide
information about the interaction of the biphoton and the metallic
film. The entanglement degradation in the focused case is
explained in the framework of a fully multimode model. This
phenomenon is a consequence of the polarization-selective
filtering behavior of the metallic nanostructured film.
\end{abstract}


\maketitle

Entanglement~\cite{sch35a,sch35b,sch35c} is one of the most
strange properties of quantum mechanics. Despite its puzzling
character, this property, which is directly linked to the
non-local nature of the theory, has been tested many times. The
first experiments were performed with simple systems comprising
two photons~\cite{fre72,asp82} or two atoms~\cite{hag97}.
Recently, entanglement has started to be considered as a resource
for diverse applications in quantum information theory. This has
driven the interest in the demonstration of entanglement for
systems involving many particles. Large systems are more prone to
decoherence processes and, therefore, entanglement should be a
very fragile property for them. For this reason the experiment of
Julsgaard \emph{et al}.~\cite{jul01} showing entanglement between
two gaseous caesium samples, and the plasmon-assisted transmission
of entangled photons shown by Altewischer \emph{et
al}.~\cite{alt02} have both attracted quite a lot of interest.
This Letter is devoted to Altewischer's experience. Although some
theoretical aspects of the experiment have been treated in the
original paper and in Ref.~\cite{vel03}, a general (multimode)
theory is still lacking. Here a complete detailed theory is
derived and a thorough analysis addressing all aspects of the
experiment is presented.

In the experiment, pairs of correlated photons are generated by
spontaneous parametric down conversion. The desired input
polarization-entangled biphoton state is obtained by appropriate
manipulation of the generated photon pairs~\cite{kwi95}. This
state is a quasi-monochromatic ($\lambda\simeq 813\,\textrm{nm}$)
quasi-plane wave (propagating along the $Z$ axis) in the
polarization singlet $(|X_1 Y_2\rangle-|Y_1 X_2\rangle)/\sqrt{2}$,
where $X$ and $Y$ denote horizontal and vertical polarization,
respectively, and the subscripts ($i=1,2$) label the first and
second photon. After travelling along their respective
trajectories parallel to the $Z$ axis, each photon traverses a
polarizer $\textrm{P}_i$ and is measured by a detector
$\textrm{D}_i$ [$\textrm{P}_i$ is orthogonal to the $Z$ axis and
rotatable around it (the angle between the optic axis of
$\textrm{P}_i$ and the $X$ axis will be denoted as $\beta_i$)].
These measured signals are electronically combined to obtain the
rate of coincident photon detections. Such a setup allows one to
determine the biphoton fringe visibility $V_{\beta_2}$ for fixed
$\beta_2$, which is a measure of the photons entanglement
degree~\cite{kwi95}.

\begin{figure}[b]
\includegraphics{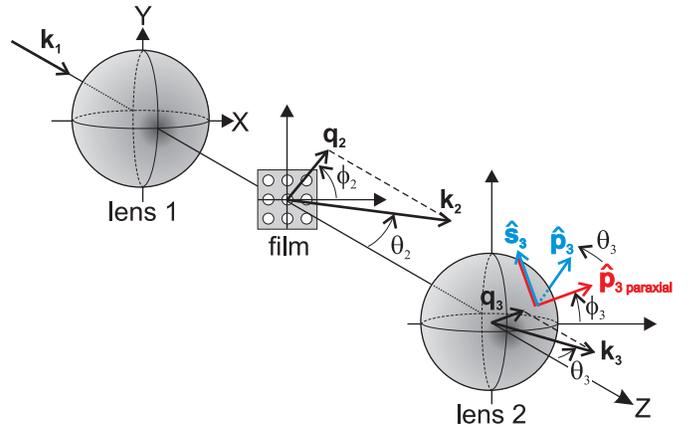}
\caption{\label{telescope}Optical elements in the trajectory of
photon 1. The lenses, that constitute a confocal telescope, are
only present in some parts of the experiment. The incident photon
is represented by a monochromatic plane wave with wave vector
$\textbf{k}_1=(\textbf{q}_1,k_{1z})=(\textbf{0},k)$. The first
lens produces a focused beam including many plane waves with wave
vectors $\textbf{k}_2=(\textbf{q}_2,k_{2z})$. The second lens
gives rise again to a bundle of modes with
$\textbf{k}_3=(\textbf{q}_3,k_{3z})$. The unit vectors
$\hat{\textbf{p}}_{3}$, $\hat{\textbf{s}}_{3}$ indicate the $p$
and $s$ polarization directions, respectively, whereas
$\hat{\textbf{p}}_{3\,\textrm{paraxial}}=\hat{\textbf{q}}_3$.}
\end{figure}

Photon 2 propagates freely from the source to $\textrm{P}_2$,
whereas photon 1 traverses a 200 nm thick Au film deposited on top
of a 0.5 mm glass substrate (Fig.~\ref{telescope}). The metallic
film is drilled with cylindrical holes (200 nm diameter) arranged
as a square lattice (period 700 nm). The transmission of photon 1
through the metallic film (at $\lambda\simeq 813\,\textrm{nm}$) is
due to the phenomenon of extraordinary light transmission mediated
by surface plasmon modes~\cite{ebb98,mar01}. The chosen wavelength
corresponds (almost) to a transmission resonance ascribed to a
surface mode at the metal-glass interface and propagating along
the diagonals of the hole array, i.e., a $(\pm 1,\pm 1)$ mode. In
order to investigate the effect of focusing on the entanglement
behavior upon transmission, the film is positioned at the focus of
a confocal telescope in some parts of the experiment (lenses'
focal length $f=15\,\textrm{mm}$, telescope's numerical aperture
0.13). Altewischer \emph{et al}. report entanglement preservation
upon transmission when no telescope is used. The entanglement is
however degraded when photon 1 is focused on the hole array and,
moreover, the measured visibilities $V_{0^{\circ}}$ and
$V_{45^{\circ}}$ are different.

A careful interaction model of the biphoton with the hole array
should explicitly include in the wave function the quantum state
of the solid. In a simplified monomode case (i.e., without
telescope), the initial wave function is
$|\Phi_{\textrm{in}}\rangle=(|X_1 Y_2\rangle-|Y_1
X_2\rangle)/\sqrt{2}\otimes|S\rangle$, where $|S\rangle$ is the
initial state of the solid. When the interaction has finished, the
wave function $|\Phi_{\textrm{out}}\rangle$ can be written as:
\begin{eqnarray}\label{quantumstate}
&  t_{X_1X_1}|X_1Y_2\rangle\otimes|S_{xx}\rangle+
   t_{Y_1X_1}|Y_1Y_2\rangle\otimes|S_{yx}\rangle\nonumber\\
-& t_{X_1Y_1}|X_1X_2\rangle\otimes|S_{xy}\rangle
  -t_{Y_1Y_1}|Y_1X_2\rangle\otimes|S_{yy}\rangle,
\end{eqnarray}
where $t_{AB}$ are the transmission amplitudes for the various
channels, and $|S_{ab}\rangle$ are normalized wave functions for
the final state of the solid. The system's final quantum state is
defined by post-selection, and for this reason
$|\Phi_{\textrm{out}}\rangle$ only includes terms with exactly two
photons. In other words, processes where, for instance, photon 1
is absorbed do indeed exist
($t_{0X_1}|0Y_2\rangle\otimes|S_{0x}\rangle-
t_{0Y_1}|0X_2\rangle\otimes|S_{0y}\rangle$), but they are not
relevant for the visibility measurement because only coincident
photons are registered. Notice that the final states of the solid
\emph{must} be taken into account, as may be clearly seen by
considering the two following extreme situations: (\emph{i})~all
$|S_{ab}\rangle$ are orthogonal to each other. In this case the
solid and the biphoton can be entangled to a larger or lesser
extent (depending on the $t_{AB}$ values) but the biphoton state
(obtained by tracing over the solid) is always a mixture of
factorizable states and it is therefore completely
disentangled~\cite{hil97}. Let us point out that in this case,
after passage of photon 1, the solid incorporates a
``which-polarization'' information linked to the photons
polarization state. This translates into an entanglement loss.
(\emph{ii})~all $|S_{ab}\rangle$ are equal. In this case the solid
and the biphoton are completely disentangled, and the biphoton
state can range from factorizable to maximally entangled depending
on the $t_{AB}$ values. Since entanglement is preserved in some
parts of Altewischer's experiment, (\emph{ii}) is our model for
the interaction process, i.e., the interaction does not introduce
``which-way'' labels in the solid. Such a theoretical framework
explains why entanglement is preserved when the photon 1 is not
focused. It is a simple consequence of two facts:  first, no
``which-way'' labels are introduced in the solid, and second, the
transfer matrix $t_{AB}$ for an orthogonally incident plane wave
(non-focused) on a square hole array is (by symmetry) proportional
to the identity.

Within the present model the visibility computation only requires
the determination of the transfer matrix $t_{AB}$ for the employed
optical set up. When the telescope is used (to focus the field at
a spot on the film), it is necessary to consider a multimode
theory. The calculation proceeds as follows: (\emph{i})~the
electromagnetic field of photon 1 is expanded in plane waves
$|\textbf{q}\,\, \sigma\rangle$ before and after each optical
element ($\sigma=\{p,s\}$ denotes the two possible polarizations;
see Fig.~\ref{telescope}). (\emph{ii})~Lens and film are described
by their transfer matrices in the aforementioned basis,
$\mathbb{L}(\textbf{q}_{2},\textbf{q}_{1})$,
$\mathbb{F}(\textbf{q}_{2})$, respectively. (\emph{iii})~These
matrices are combined to obtain the transfer matrix
$\mathbb{T}(\textbf{q}_{3},\textbf{q}_{1})$ of the telescope with
the hole array inside it. (\emph{iv})~The biphoton transfer matrix
$\mathbb{M}(\textbf{q}_{3},\textbf{q}_{1})$ for the whole set up
(including polarizers) is then given by the tensor product of the
transfer matrices for each photon, i.e.,
$\mathbb{M}(\textbf{q}_{3},\textbf{q}_{1})=\mathbb{P}_1(\textbf{q}_{3})\mathbb{T}(\textbf{q}_{3},\textbf{q}_{1})\otimes\mathbb{P}_2(\textbf{0})$,
where $\mathbb{P}_i$ are the polarizers' transfer matrices. In the
case of normal incidence, the telescope plus hole array transfer
matrix $\mathbb{T}(\textbf{q}_{3},\textbf{0})$ turns out to be:
\begin{equation}\label{telescopematrix}
\mathbb{R}(\phi_3) \int d\textbf{q}_2 e^{i\frac{(n-1)\Delta}{2 n
k}(\textbf{q}_2-\frac{n f}{(n-1)\Delta}\textbf{q}_3)^2}
\mathbb{R}^{-1}(\phi_2)\mathbb{F}(\textbf{q}_2)\mathbb{R}(\phi_2),
\end{equation}
where $\mathbb{R}$ is the two-dimensional rotation matrix, $n$ and
$\Delta$ are the substrate refractive index and thickness,
respectively ($n=1.52$), $k$ is the wave number, $f$ is the focal
length, and the remaining variables are explained in
Fig.~\ref{telescope}. Note that the rapidly oscillating phase
inside the integral means that the amplitude of
$|\textbf{q}_3\rangle$ is mainly given by the
$|\textbf{q}_2\rangle$ modes inside the telescope around the
stationary phase condition $\textbf{q}_2=n f
\textbf{q}_3/(n-1)\Delta$.

\begin{figure}[h]
\includegraphics{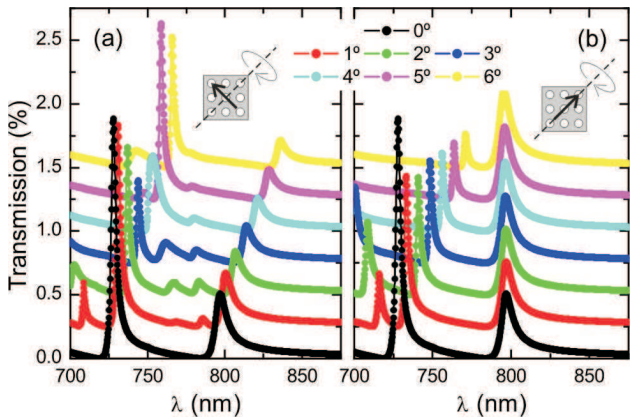}
\caption{\label{transmissions}Hole array transmittance as a
function of wavelength. The film is illuminated by an orthogonally
incident plane wave (non-focused), and it is tilted (from
$0^{\circ}$ to $6^{\circ}$) around the diagonal. The incident
field linear polarization is perpendicular to the lattice diagonal
in (a) and parallel to it in (b). For clarity the curves are
offset 0.25\%.}
\end{figure}

To obtain Eq.~(\ref{telescopematrix}) a few approximations have
been done. First, due to the low numerical aperture of the
telescope, paraxial equations can be used. The lens transfer
matrix $\mathbb{L}(\textbf{q}_{2},\textbf{q}_{1})$ is given by:
\begin{equation}\label{lensmatrix}
\frac{f}{2\pi k
i}e^{i\frac{f}{2k}(\textbf{q}_2-\textbf{q}_1)^2}\mathbb{R}(\phi_2)\mathbb{R}^{-1}(\phi_1),
\end{equation}
where the subscripts ($j=1,2$) refer to modes before and after the
lens, respectively. Free propagation of a $|\textbf{q}_2\,\,
\sigma_2\rangle$ mode inside the telescope along a distance $z$
amounts to an extra phase that, apart from global factors, is
given by $\exp(-i z \textbf{q}_2^2/2 k)$ in the paraxial
approximation. Second, the detectors are located at the back focal
plane of an auxiliary lens placed after the polarizers. This
implies that each point in the detector essentially collects one
$|\textbf{q}_3\rangle$ (in the limit of very large auxiliary lens
aperture). For this reason the relative phases of the
$|\textbf{q}_3\rangle$ modes after the telescope are not relevant
for the visibility computation. Third, concerning the hole array,
again due to the low numerical aperture of the telescope, it is
enough to keep the 0-th diffracted order (higher orders are not
collected by the second lens), and therefore the film transfer
matrix $\mathbb{F}(\textbf{q})$ is $\textbf{q}$-diagonal. This
matrix is numerically computed as explained in~\cite{mar01}.
Figure~\ref{transmissions} shows the computed transmittance when
the hole array is illuminated by an orthogonally incident
(non-focused) plane wave, and the film is tilted around the
diagonal (compare to Figs. 1b, 1c in~\cite{alt02}). Despite the
simulations do not exactly give the experimental peaks' heights
and widths values, all main features are reproduced, including
number and position of peaks, and overall order of magnitude. The
behavior of the photonic bands as a function of parallel momentum
is also correctly described.

The transfer matrix $\mathbb{T}(\textbf{q}_{3},\textbf{0})$
corresponding to telescope plus film can be worked out
analytically when $\textbf{q}_3=\textbf{0}$, i.e., when the
detector's aperture is extremely small. In this case the matrix
before the integral in Eq.~(\ref{telescopematrix}) disappears and
the phase inside the integrand does not depend on $\phi_2$. If one
writes down explicitly $\mathbb{T}(\textbf{0},\textbf{0})$ and
takes into account the symmetry properties of the hole array, it
can be shown that $\mathbb{T}(\textbf{0},\textbf{0})$ is
proportional to the identity [note that this result does not
depend on the particular model for the numerical computation of
$\mathbb{F}(\textbf{q})$]. For this reason the whole set up should
again preserve entanglement when only the following channels
$(\textbf{q}_1=\textbf{0})\rightarrow (\textrm{all}\,\,
\textbf{q}_2) \rightarrow (\textbf{q}_3=\textbf{0})$ are
considered. This means that the focusing of photon 1 does not by
itself degrade entanglement in every situation but, rather, only
when the transfer matrix $t_{AB}$ is not proportional to the
identity. This could be easily checked in an experiment by
inserting an iris before the detector.

\begin{figure}[h]
\includegraphics{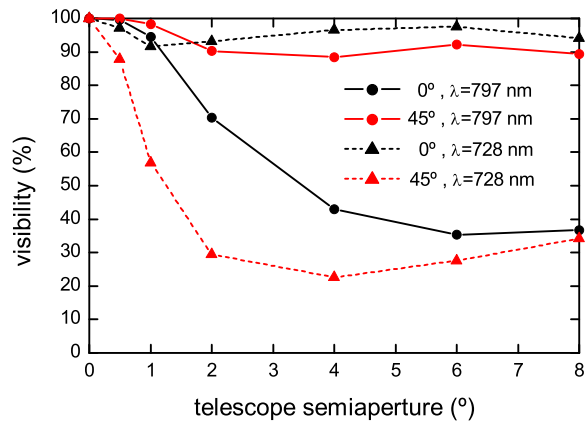}
\caption{\label{visibilities}Biphoton fringe visibility as a
function of the telescope semiaperture. The visibilities are shown
for two orientations of the second polarizer: $\beta_2 =
0^{\circ},\, 45^{\circ}$. The chosen wavelengths are the
resonances shown in Fig.~\ref{transmissions} for no tilt: 797 nm
(corresponding to a surface mode propagating along the array
diagonals), and 728 nm (corresponding to a mode propagating along
the $X$ or $Y$ axes).}
\end{figure}

Figure~\ref{visibilities} shows the visibilities obtained when the
telescope is again employed, but now \emph{all} channels
[$(\textbf{q}_1=\textbf{0})\rightarrow (\textrm{all}\,\,
\textbf{q}_2) \rightarrow (\textrm{all}\,\, \textbf{q}_3)$] are
considered. The visibility decreases as the telescope semiaperture
increases (for $0^{\circ}$ semiaperture the monomode case is
obviously recovered). The same behavior observed in~\cite{alt02}
for the $(\pm 1,\pm 1)$ mode is reproduced by the simulations for
$\lambda = 797\, \textrm{nm}$ and telescope semiaperture
$8^{\circ}$: $V_{45^{\circ}}=89\%$, $V_{0^{\circ}}=37\%$ (87 and
73 in the experiment, respectively). The numerical discrepancy for
$V_{0^{\circ}}$ can be essentially attributed to the fact that the
computed transmission resonances are narrower than the measured
ones, and the visibility is very sensitive to this variable. It is
also to be noted that in~\cite{alt02} the employed wavelength is a
bit larger than the resonant wavelength whereas we have computed
the visibility at the resonance maximum itself ($\lambda = 797\,
\textrm{nm}$). Our simulations show that $V_{0^{\circ}}$ grows for
wavelengths larger than the resonant one ($V_{0^{\circ}}=52\%$ for
$\lambda = 802\, \textrm{nm}$), whereas $V_{45^{\circ}}$ remains
approximately  the same. Note that the visibilities are not
monotone functions of the semiaperture. This is due to the fact
that for larger semiapertures, higher values of $\theta_2$ are
included in the integral of Eq.~(\ref{telescopematrix}), and this
permits the excitation of surface modes different from $(\pm 1,\pm
1)$ (as can be indirectly seen in Fig.~\ref{transmissions}).
\begin{figure*}
\includegraphics{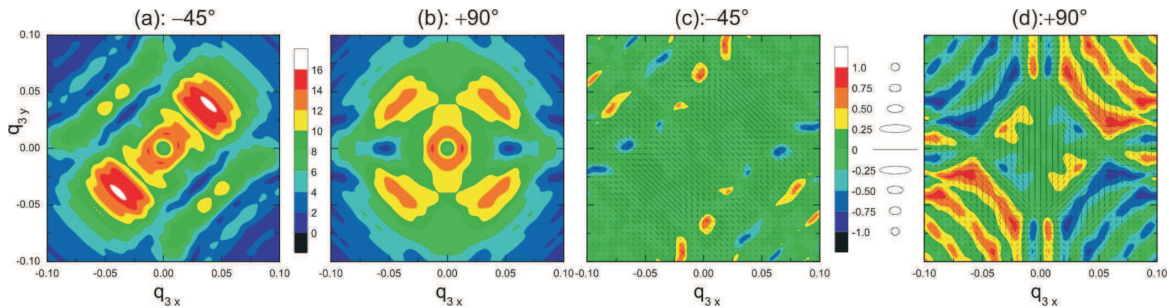}
\caption{\label{polarization}Intensities (a,~b) and polarizations
(c,~d) of mode $|\textbf{q}_3\rangle$ after the telescope as a
function of $(q_{3x},q_{3y})$. The excitation is one single
linearly polarized photon ($\lambda = 797\, \textrm{nm}$). The
incident photon polarizations are $-45^{\circ}$ in (a,~c) and
$90^{\circ}$ in (b,~d). The telescope semiaperture is $8^{\circ}$.
The left color scale (arb. units) is shared by (a) and (b), and
the right color scale corresponds to (c) and (d). In (c,~d) the
color scale represents the ratio of the minor to the major
polarization elipse axes (for instance, $\pm1$ correspond to left
and right circular polarization), and the small line segments
indicate the orientation of the elipse major axis. The maxima (and
minima) of all abscissa and ordinate scales correspond to a
semiaperture of $\theta_3=0.1^{\circ}$.}
\end{figure*}

The visibility reduction (as compared to the non-focused case) can
be understood because the transfer matrix
$\mathbb{T}(\textbf{q}_{3},\textbf{0})$ of the telescope plus film
is not proportional to the identity anymore. This means that the
system acts as a polarization-selective filter. The initial
balance between the $|X_1 Y_2\rangle$ and $|Y_1 X_2\rangle$
components (which is responsible of the maximal entanglement of
the input state) is therefore destroyed and as a consequence the
entanglement is degraded. In the following it will be explained
why do $V_{45^{\circ}}$ and $V_{0^{\circ}}$ behave differently.

Let us remind that, when $V_{\beta_2}$ is measured, polarizer
$\textrm{P}_2$ is set with this $\beta_2$. To understand the
behavior of $V_{\beta_2}$ one can plot the field after the
telescope due to a single photon 1 incident with
$\beta_2\!+\!90^{\circ}$ linear polarization. This is so because
the singlet biphoton state can be written as $(|\beta_2\,\,
\beta_2\!+\!90^{\circ}\rangle-|\beta_2\!+\!90^{\circ}\,\,
\beta_2\rangle)/\sqrt{2}$ for every $\beta_2$ (i.e., it is
polarization isotropic). $V_{\beta_2}$ will be 100\% if there
exists an orientation $\beta_1$ for polarizer $\textrm{P}_1$ that
completely blocks the field due to photon 1 incident with linear
polarization $\beta_2\!+\!90^{\circ}$. Otherwise, as $\beta_1$ is
varied, the collected intensity oscillates between a non-zero
minimum and a maximum, and $V_{\beta_2}<100\%$. Such a
polarization information is displayed in Fig.~\ref{polarization}.
Let us start with the analysis for $V_{45^{\circ}}$. For
$-45^{\circ}$ incident polarization of photon 1, only the
$(+1,-1)$ and $(-1,+1)$ modes are excited. The output field is
predominantly linearly polarized along $-45^{\circ}$
[Fig.~\ref{polarization}(c)], yielding high $V_{45^{\circ}}$ (a
low photon coincidence  rate will be registered for
$\beta_1=+45^{\circ}$). On the other hand the analysis for
$V_{0^{\circ}}$ goes as follows. For $90^{\circ}$ photon 1
incident polarization, all $(\pm1,\pm1)$ modes are excited, the
output field is generally eliptically polarized
[Fig.~\ref{polarization}(d)], and it includes various polarization
directions. This fact is responsible for the decrease in
$V_{0^{\circ}}$ (no $\beta_1$ exists that nearly blocks the
field). Note that the central region of both diagrams is linearly
polarized along the incident polarization direction. This fits
with the transfer matrix being approximately proportional to the
identity (for this region) and with the 100\% visibility expected
for small detector aperture, as discussed previously. It is also
to be noticed that Fig.~\ref{polarization}~(a) and (b) compare
well with Figs. 2b, and 2d in~\cite{alt02b} [the stationary phase
condition mentioned above maps the diagram's side lengths shown
here ($2\theta_3=0.2^{\circ}$) to the $2\theta_2=17.6^{\circ}$
value shown in~\cite{alt02b}; as opposed to~\cite{alt02b},
circular fringes do not appear in Fig.~\ref{polarization}~(a) and
(b) in the range shown because multiple interferences in the
substrate were not included in the calculation]. The $\lambda =
728\, \textrm{nm}$ surface mode propagates along the $(\pm1,0)$
and $(0,\pm1)$ directions and, as a consequence of previous
discussion, the roles of $V_{45^{\circ}}$ and $V_{0^{\circ}}$
should be exchanged (this is indeed observed in the polarization
diagrams for $\lambda = 728\, \textrm{nm}$, not shown here for
brevity). In fact, compared to $\lambda = 797\, \textrm{nm}$, the
high and low visibility values are now exchanged, as it is
distinctly seen in Fig.~\ref{visibilities}, where $V_{45^{\circ}}$
is low and $V_{0^{\circ}}$ is high.

In conclusion, a detailed multimode theoretical analysis of
Altewischer \emph{et al.} experiment has been presented.
Entanglement preservation in the monomode case implies a
particular model for the hole array-biphoton interaction, namely,
this interaction cannot introduce ``which-way'' labels in the
metallic film. Our model also reproduces the measured results in
the focused case. The entanglement degradation is understood as a
consequence of the polarization-selective filtering behavior of
the hole array for non-orthogonal incidence. A polarization
analysis explains the different values of the $0^{\circ}$ and
$45^{\circ}$ visibilities.


\end{document}